\documentclass[aps,amsmath,twocolumn,showpacs]{revtex4}
\usepackage{graphicx}
\usepackage{mathrsfs}
\usepackage{bm}% bold math
\usepackage{amssymb}
\usepackage{amsmath}
\usepackage{textcomp}
\usepackage{url}
\usepackage{dsfont}	
\usepackage{braket}
\usepackage[colorlinks=true,linkcolor=magenta,citecolor=blue]{hyperref}%

\usepackage{empheq}
\usepackage{multirow}
\begin{document}

\preprint{APS/123-QED}

\title{Controllable magnon-induced transparency in a ferromagnetic material via cross- and self-Kerr effects}
\author{Akhtar Munir}
\email{amkhan@mail.ustc.edu.cn}
\affiliation{ Zhejiang University-University of Illinois at Urbana-Champaign (ZJU-UIUC) Institute, Zhejiang University, Haining, Zhejiang, China}
%
%\date{}
%


\begin{abstract}
Nonlinear interactions between optical fields and magnetic modes in cavity magnonics constitute a rich source of various nontrivial  effects in optics and quantum information processing. In cavity magnonics, the nonlinear cross-Kerr effect, which shifts the cavity's central frequency when a magnetic material is pumped, causes the system to exhibit both Kittle and magnetostatic modes. Here, we propose a new scheme for the investigation of probe fields transmission profiles in cavity magnonic systems composed of a microwave cavity and a ferromagnetic material (Yttrium iron garnet sphere). We report single-to-double magnon-induced transparency (MIT) dips and a sharp magnon-induced absorption (MIA) peak, and demonstrate how nonlinear cross- and self-Kerr interactions can significantly enhance or suppress these phenomena. It is observed that the splitting of the MIT window occurs when we incorporate magnon-magnon modes coupling, which helps introducing a new degree of freedom to light-matter interaction problems. Moreover, we investigate the propagation of group delay in the vicinity of transparency and demonstrate how a sharp dip allows the realization of slow light for a longer period of time. We found that both the cavity-Kittle and magnon-magnon modes coupling parameters influence the propagation of group delay, which demonstrates how subluminal-to-superluminal (and vice versa) propagation phenomena may occur and transform. These findings could pave the way for future research into nonlinear effects with novel applications in cavity magnonics devices, which might be exploited for several applications such as quantum computing devices and quantum memories.
%\todo{use this command in LaTeX file for your suggestion}
\end{abstract}

\maketitle

%%%%%%%%%%%%%%%%%%%%%%%%%%  body  %%%%%%%%%%%%%%%%%%%%%%%%%%
\section{Introduction}
\label{Sec:intro}
Over the last few years, cavity magnonics has increasingly demonstrated significant advantages in fundamental and applied research~\cite{Zhang2016,Zhang2014,Bai2015,Zhang2015}, and it is anticipated to be a vital part of hybrid quantum systems~\cite{Tabuchi2019} and quantum network nodes~\cite{Wang2021}. The most common physical realization of a cavity magnonic system is a microwave cavity and ferromagnetic material, such as a yttrium iron garnet (YIG) sphere, which has drawn extensive research and experienced amazing performance in recent decades. The principal reason is that YIGs have relatively high spin densities ($\sim 4.22 \times 1027$ m$^{-3}$) and low damping rates ($\sim 1$ MHz), which are necessary for generating strong coupling~\cite{Tabuchi2014,Goryachev2014} between the microwave cavity photon mode and the YIG's magnon mode, which enables quantum information transfer. So far, insights from experimental and theoretical research based on cavity magnonics have revealed a number of phenomena, including magnon dark modes~\cite{Zou2015}, magnon-induced transparency (MIT)~\cite{Wang2018}, entanglement which is resource of quantum technologies including quantum computing~\cite{Raussendorf2001,Knill2001}, quantum teleportation~\cite{Pirandola2015} and quantum metrology~\cite{Giovannetti2006,Giovannetti2011}, non-Hermitian physics~\cite{Wang2019,Yang2020,Harder2017} and nonclassical states~\cite{Li2018,Li2019,Zhang2019}. 

 In hybrid quantum systems, light-matter interaction is induced by nonlinearities such as self-Kerr~\cite{Zhang2019,Yang2021} and cross-Kerr~\cite{Yang2022} effects in cavity magnonic systems, radiation pressure interactions in optomechanical systems~\cite{Vitali2007,Genes2008}, magnetostrictive interactions in cavity magnomechanical systems~\cite{Zhu2019}, and other systems involving parametric amplifiers~\cite{Nair2020,Liu2021}. These nonlinear interactions are weak and challenging to detect in certain systems, but they can become strong enough to be the dominant factor in others. The cross-Kerr effect, for example, is one of the complex nonlinear interactions between fields and waves that can occur in superconducting circuits~\cite{Hoi2013,Vrajitoarea2020}, natural ions~\cite{Ding2017} and atoms~\cite{He2014,Xia2018}. The cross-Kerr effect is a nonlinear change in the frequency of a resonator as a function of the number of excitations in another mode that engage the resonator. Another nonlinear effect arising from magnetocrystalline anisotropy in a YIG sample is self-Kerr nonlinearity~\cite{Zhang2019,Yang2021}, which is typically weak~\cite{Shen2021} but can be amplified by driving the corresponding spin-wave modes with a drive field. Thus, having an understanding of these nonlinear interactions is not only of fundamental significance, but also useful in a number of different applications. For example, the cross-Kerr effect can be used to construct quantum logic gates~\cite{Turchette1995,Brod2016}, perform quantum non-demolition measurements~\cite{Miranowicz2019,Dassonneville2020} and to generate entangled photons~\cite{Sheng2008}.

Motivated by new advancements in hybrid magnomechanical systems, we construct a cavity magnonics system composed of a ferromagnetic material that supports both Kittle and magnetostatic (MS) modes. The goal is to investigate the consequences of nonlinear cross- and self-Kerr interaction on the MIT (MIA) phenomenon caused by destructive (constructive) interference of optical fields inside the cavity, where the magnon Kerr effect is caused by magnetocrystalline anisotropy in the YIG sphere. We observe a single-MIT window due to cavity field interaction with the Kittle mode, which then split into two windows when the Kittle mode interacts with the MS mode, which incorporated another degree of freedom. We explore the self-Kerr effect of both spin modes, specifically the Kittle and MS modes, and notice that the self-Kerr effect of the Kittle mode is responsible for the asymmetric behavior of the MIA profile, whereas the self-Kerr effect of the MS mode splits the single-MIT window into two. Furthermore, the slow and fast light effects are examined in the vicinity of two MIT windows for various control parameter values, namely the cavity-Kittle mode and cross-Kerr coupling parameter. We report both slow and fast light effects in a single setup, which is an advantage of the proposed scheme over previous work that only demonstrated slow~\cite{Chen2011,Zhan2013,Jiang2013} or fast~\cite{Tarhan2013} light propagation.

The article is arranged as follows. In Sec.~\ref{Sec:model}, we present a theoretical model of a general cavity magnonic system consisting a YIG sphere and introduce the effective Hamiltonian for the proposed system. To investigate the dynamics of the system, the quantum Langevin equations are derived then used to deduce a mathematical formula for the outgoing probe field is obtained by employing the standard input-output method. In Sec.~\ref{Sec:results}, numerical results are provided to illustrate the realization and control of single-to-double MIT windows profiles, investigate the impact of cross- and self-Kerr effects on the MIT phenomenon, and to demonstrate a mechanism for the switching from slow to fast light. Finally, we end our work with conclusions in Sec.~\ref{Sec:conclusion}. 

\begin{figure*}
	\includegraphics[width=0.7\linewidth]{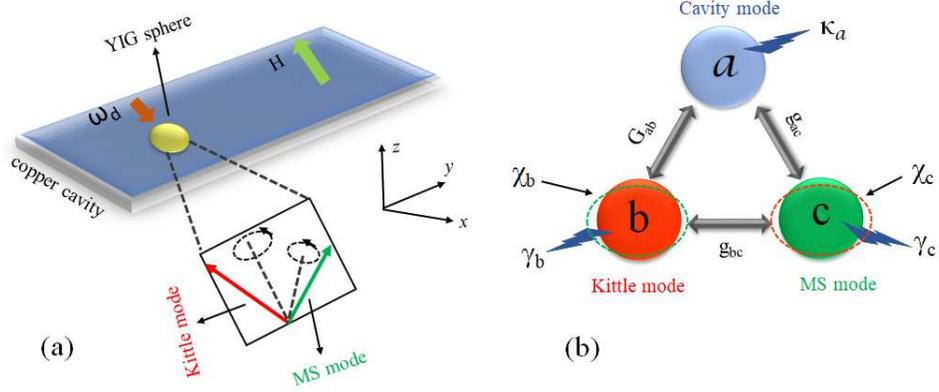}
\caption{(a) Sketch of a cavity magnonics system consisting of a YIG sphere mounted on a copper cavity and placed in a bias magnetic field polarized along the $z$ direction. Here, $\omega_d$ is the driving field frequency. The Kittle and magnetostatic (MS) modes are the two spin-wave modes of the YIG sphere. (b) Schematic diagram showing the coupling of cavity mode with Kittle and MS mode, where g$_{ab}$, g$_{ac}$, g$_{bc}$ are the coupling parameters. $\mathcal{X}_b$ and  $\mathcal{X}_c$ are the Kittle and MS modes self-Kerr coefficients, while $\kappa_a$, $\gamma_b$, $\gamma_c$ represent the dissipation rates associated with cavity, Kittle and MS modes, respectively. }
	\label{fig:one}
\end{figure*}

\section{Model and equation of motion}
\label{Sec:model}
We consider the cavity magnonic system illustrated in Fig. \ref{fig:one}. The ferromagnetic material (YIG sphere) is positioned within a microcavity which contains dispersive spin waves in which the spatially uniform Kittle mode interacts strongly with cavity photons. As compared to the Kittle mode, the MS mode with a finite wave number has a distinct frequency that can be used to execute selective excitation through driving field wavelength and cavity mode selection~\cite{Zhang2019}. After using the dipole and rotating wave approximations, the effective Hamiltonian of the proposed cavity magnonic system can be written as
\begin{equation}\label{Hamiltonian}
\begin{aligned}
\hat{H}_\text{eff}=& \Delta_{a} \hat{a}^{\dagger} \hat{a}+\Delta_{b} \hat{b}^{\dagger} \hat{b}+\Delta_{c} \hat{c}^{\dagger} \hat{c}+\mathcal{X}_{b} \hat{b}^{\dagger} \hat{b} \hat{b}^{\dagger} \hat{b}+\mathcal{X}_{c} \hat{c}^{\dagger} \hat{c} \hat{c}^{\dagger} \hat{c} \\
&+G_{ab}\left(\hat{a}^{\dagger} \hat{b}+\hat{a} \hat{b}^{\dagger}\right)+\text{g}_{bc}\hat{b}^{\dagger} \hat{b} \hat{c}^{\dagger} \hat{c}+\text{i}\Omega_{b}\left(\hat{b}^{\dagger}-\hat{b}\right) \\
&+\text{i}\Omega_{c}\left(\hat{c}^{\dagger}-\hat{c}\right)+\text{i}\mathcal{E}_{p}\left(\hat{a}^{\dagger}\text{e}^{-\text{i}\Delta_p t}-\hat{a}\text{e}^{\text{i}\Delta_p} t\right),
\end{aligned}
\end{equation}
where $\hbar=1$. The first three terms in Eq. \eqref{Hamiltonian} describe the free Hamiltonian of the cavity, the Kittle and the MS modes, respectively; here, $\hat{a}^\dagger$($\hat{a}$), $\hat{b}^\dagger$($\hat{b}$) and $\hat{c}^\dagger$($\hat{c}$) are their creation (annihilation) operators, and $\Delta_{a,b,c}=\omega_{a,b,c}-\omega_d$ define the corresponding detunings. Here, $\mathcal{X}_{b}$ and $\mathcal{X}_{c}$ are the self-Kerr coefficients of the Kittle and MS modes, respectively, while $G_{ab}$ represents the coupling strength between the cavity mode and the Kittle mode. On the other hand, the quantity $\text{g}_{bc}$ captures the magnon-magnon mode, also known as cross-Kerr coefficient parameter. We exclude the coupling between the cavity mode and the MS mode because, on a micrometer-scale YIG sphere, the spin moment of the Kittel mode contributes more to the dipole than that of the MS mode~\cite{Wu2021}.
The last three terms denote driving field interactions with the Kittle and MS modes, with Rabi frequencies $\Omega_{b}$ and $\Omega_{c}$, respectively, and prob field interaction with the cavity mode, with field strength of $\mathcal{E}_{p}$ and detuning $ \Delta_p=\omega_{p}-\omega_{d}$.

To investigate the dynamics of the proposed system, we deploy the Heisenberg-Langevin approach~\cite{Walls1994} in order to derive the following system of operator coupled differential equations, the well-known quantum Langevin equations (QLEs):
\begin{equation}
\label{eq.QLEs1}
\begin{aligned}
\frac{\text{d} \hat{a}}{\text{d} t}=&-\left(\kappa_{a}+\text{i} \Delta_{a}\right) \hat{a}-\text{i} G_{ab} \hat{b}+\mathcal{E}_{p}\text{e}^{-\text{i}\Delta_p t}+\sqrt{2 \kappa_{a}} \hat{a}^{\mathrm{in}}, \\
\frac{\text{d} \hat{b}}{\text{d} t}=&-\left(\gamma_{b}+\text{i} \Delta_{b}\right) \hat{b}-\text{i} G_{ab} \hat{a}+ \Omega_{b}-\text{i} \text{g}_{bc} \hat{b} \hat{c}^{\dagger} \hat{c} -2 \text{i} \mathcal{X}_{b} \hat{b}^{\dagger} \hat{b} \hat{b}\\&+\sqrt{2 \gamma_{b}} \hat{b}^{\mathrm{in}}, \\
\frac{\text{d} \hat{c}}{\text{d} t}=&-\left(\gamma_{c}+\text{i}\Delta_{c}\right) \hat{c}-2 \text{i} \mathcal{X}_{c} \hat{c}^{\dagger} \hat{c} \hat{c}-\text{i} \text{g}_{bc} \hat{c} \hat{b}^{\dagger} \hat{b}+\Omega_{c} +\sqrt{2 \gamma_{c}} \hat{c}^{\mathrm{in}},
\end{aligned}
\end{equation}
where $\kappa_a$, $\gamma_{b}$ and $\gamma_{c}$ represent the damping rates of the dissipation processes associated with modes $a$, $b$, and $c$, respectively. Following the standard quantum Langevin approach, noise operators $\hat{a}^{\mathrm{in}}$, $\hat{b}^{\mathrm{in}}$ and $\hat{c}^{\mathrm{in}}$ for the cavity, the Kittle and MS modes, respectively, are introduced; these are input operators obeying certain generic properties, such as having zero-mean fluctuations with statistics adhering to correlation functions of the following type~\cite{Gardiner2000}: 
\begin{equation}
    \begin{aligned}
       \left\langle \hat{a}_{\text {in }}(t) \hat{a}_{\text {in}}^{\dagger}\left(t^{\prime}\right)\right\rangle=&\delta\left(t-t^{\prime}\right),\\
       \left\langle \hat{a}_{\text {in }}(t) \hat{a}_{\text {in}}^{\dagger}\left(t^{\prime}\right)\right\rangle=&\delta\left(t-t^{\prime}\right), \\ 
       \left\langle \hat{a}_{\mathrm{in}}^{\dagger}(t) \hat{a}_{\mathrm{in}}\left(t^{\prime}\right)\right\rangle=&0, \\[4pt] 
       \left\langle \hat{o}_{\mathrm{in}}^{\dagger}(t) \hat{o}_{\mathrm{in}}\left(t^{\prime}\right)\right\rangle=&n_{\mathrm{th}} \delta\left(t-t^{\prime}\right),\\
       \left\langle \hat{o}_{\mathrm{in}}(t) \hat{o}_{\mathrm{in}}^{\dagger}\left(t^{\prime}\right)\right\rangle=&\left(n_{\mathrm{th}}+1\right) \delta\left(t-t^{\prime}\right),
    \end{aligned}
\end{equation}
where $o:=\{b,c\}$, $k_{B}$ is the Boltzmann constant, and $T$ the thermodynamic bath's temperature. In particular, $n_{\mathrm{th}}=\left[\exp{(\hbar \omega_{m} / k_{B} T)}-1\right]^{-1}$ is the average thermal photon number. 

Since the microwave drive field interacts strongly with the microcavity, a beam-splitter-like type of interaction, which couples magnons with the optomechanical system, will result in modal field behaviour characterized by large amplitudes in both the magnon and cavity fields cases. In other words, we have $|\left\langle a \right\rangle|\gg 1$, $|\left\langle b \right\rangle|\gg 1$, and $|\left\langle c \right\rangle|\gg 1$. Therefore, we can apply the standard linearization approach~\cite{Walls1994} to Eq.~(\ref{eq.QLEs1}) by expanding each operator as $\hat{\mathcal{O}}=\langle \mathcal{O}\rangle+\delta \mathcal{O}$, where $\mathcal{O}:=\{a,b,c\}$. Here, the small operator $\delta O$ captures the first-order (linear) perturbation process, which quantifies how near the system is to the thermodynamic steady state after interaction with the external drive field. Such decomposition tacitly assumes that all higher-order fluctuations processes may be neglected. Building on the linearization scheme described above, a new set of simplified differential equations can be obtained from Eq.(\ref{eq.QLEs1}) written as follows:
\begin{equation}
\label{eq.QLEs2}
\begin{aligned}
\frac{\text{d} \langle a\rangle}{\text{d} t}=&-\left(\kappa_{a}+\text{i} \Delta_{a}\right) \langle a\rangle-\text{i} G_{ab} \langle b\rangle+\mathcal{E}_{p}\text{e}^{-\text{i}\Delta_p t}, \\[4pt]
\frac{\text{d} \langle b\rangle}{\text{d} t}=&-\left(\gamma_{b}+\text{i} \Delta_{b}\right) \langle b\rangle-\text{i} G_{ab} \langle a\rangle-\text{i} \text{g}_{bc} |\langle c\rangle|^2 \langle b\rangle\\& -2 \text{i} \mathcal{X}_{b} |\langle b\rangle|^2 \langle b\rangle + \Omega_{b}, \\[4pt]
\frac{\text{d} \langle c\rangle}{\text{d} t}=&-\left(\gamma_{c}+\text{i}\Delta_{c}\right) \langle c\rangle-2 \text{i} \mathcal{X}_{c} |\langle c\rangle|^2 \langle c\rangle-\text{i} \text{g}_{bc} |\langle b\rangle|^2 \langle c\rangle+\Omega_{c}.
\end{aligned}
\end{equation}
The steady-state solutions $a_s, b_s, c_s$ of the linearized system \eqref{eq.QLEs2} may now be expressed as
\begin{equation}
    \begin{aligned}
        a_s=&\frac{-\text{i}G_{ab}b_s}{(\kappa_a +\text{i}\Delta_{a})},  \ \
        b_s=&\frac{-\text{i}G_{ab}a_s+\Omega_{b}}{(\gamma_b +\text{i}\Delta_{b}^\prime)},\  \
        c_s=&\frac{\Omega_{c}}{(\gamma_c +\text{i}\Delta_{c}^\prime)},
    \end{aligned}
\end{equation}
where
\begin{equation}
    \begin{aligned}
    \Delta_{b}^\prime:=&\Delta_{b}+2\text{i}\mathcal{X}_b |\langle b\rangle|^2+\text{g}_{bc}|\langle c\rangle|^2, \\[4pt]
    \Delta_{c}^\prime:=&\Delta_{c}+2\text{i}\mathcal{X}_c |\langle c\rangle|^2 +\text{g}_{bc}|\langle b\rangle|^2,
    \end{aligned}
\end{equation}
are the effective magnon-mode drive field detunings. On the other hand, the quantum fluctuations $\delta a, \delta b, \delta c$ themselves obey the following reduced QLEs:
\begin{equation}
\label{eq.QLEs3}
    \begin{aligned}
    \delta \dot{a}=& -(\kappa_{a}-\text{i}\Delta_{a})\delta a-\text{i}G_{ab}\delta b+\mathcal{E}_{p}\text{e}^{-\text{i}\Delta_p t},\\[4pt]
    \delta \dot{b}=& -(\gamma_{b}-\text{i}\Delta_{b}^\prime)\delta b -\text{i}\mathcal{X}_b^\prime \delta_b^\dagger -\text{i}G_{ab}\delta a+\Omega_{b}-\text{i}G_{bc}(\delta c^\dagger+\delta c),\\[4pt]
    \delta \dot{c}=& -(\gamma_{c}-\text{i}\Delta_{c}^\prime)\delta c -\text{i}\mathcal{X}_c^\prime \delta c^\dagger -\text{i}G_{bc}(\delta b^\dagger+\delta b)+\Omega_{c},
    \end{aligned}
\end{equation}
where $\mathcal{X}_b^\prime:=2\mathcal{X}_b \langle b\rangle^2$ and $\mathcal{X}_c^\prime:=2\mathcal{X}_c \langle c\rangle^2$ are the effective self-Kerr coefficients, while $G_{bc}=\langle b\rangle \langle c\rangle$ represents the effective magnon-magnon coupling strength. In addition, we also introduce
\begin{equation}
    \begin{aligned}
     \Delta_{b}^\prime=&\Delta_{b}+2\mathcal{X}_b^\prime+G_{bc}\langle c\rangle^2, \\[4pt]
     \Delta_{c}^\prime=&\Delta_{c}+2\mathcal{X}_c^\prime+G_{bc}\langle b\rangle^2,
    \end{aligned}
\end{equation}
which are the effective magnon-mode drive field detunings of the Kittle and MS modes, respectively. 
In order to solve the differential equation system \eqref{eq.QLEs3}, we make use of the following ansatz~\cite{Fabre1994,Arcizet2006}:
\begin{equation}
\label{eq.ansatz}
\begin{aligned}
\delta a &=a_{-} \text{e}^{-\text{i} \Delta_p t}+a_{+} \text{e}^{\text{i} \Delta_p t}, \\[4pt]
\delta b &=b_{-} \text{e}^{-\text{i} \Delta_p t}+b_{+} \text{e}^{\text{i} \Delta_p t}, \\[4pt]
\delta c &=c_{-} \text{e}^{-\text{i} \Delta_p t}+c_{+} \text{e}^{\text{i} \Delta_p t},
\end{aligned}
\end{equation}
where $a_\pm$, $b_\pm$ and $c_\pm$ are the fluctuation operators of the cavity, Kittle and MS modes, respectively. After substituting Eq.~(\ref{eq.ansatz}) into Eq.~(\ref{eq.QLEs3}), performing some manipulations, we arrive at the following relation for $a_-$:
\begin{widetext}
\begin{equation}
a_{-}=\frac{\mathcal{M}}{\mathcal{N}},
\end{equation}
where
\begin{align*}
\mathcal{M}=& \alpha_{1}\mathcal{E}_p \left[(1-\mathcal{X}_b^{\prime 2}\alpha_{2}\alpha_{2}^*+G_{ab}^2\alpha_{1}^*(1-\text{i}\mathcal{X}_b^\prime \alpha_{2})\alpha_{2}^*)(-1+\mathcal{X}_c^{\prime 2}\alpha_{3}\alpha_{3}^*)+G_{bc}^2\{\text{i}\alpha_{2}^*+\alpha_{2}(-\text{i}+2\mathcal{X}_b^\prime\alpha_{2}^*)\}\right.\\&+\left.\{-\text{i}\alpha_{3}+\alpha_{3}^*(\text{i}+2\mathcal{X}_c^\prime \alpha_{3})\}\right],\\[4pt]
\mathcal{N}=& \left[G_{bc}^2(-\text{i}\alpha_{2}+\text{i}\alpha_{2}^*+2\mathcal{X}_{b}^\prime \alpha_{2}\alpha_{2}^*)(\text{i}\alpha_{3}^*-\text{i}\alpha_{3}+2\mathcal{X}_{c}^\prime \alpha_{3}\alpha_{3}^*)+G_{ab}^4\alpha_{1}\alpha_{1}^*\alpha_{2}\alpha_{2}^*(-1+\mathcal{X}_{c}^2\alpha_{3}\alpha_{3}^*)\right.\\&\left.-(-1+\mathcal{X}_{b}^2\alpha_{2}\alpha_{2}^*)(-1+\mathcal{X}_{c}^2\alpha_{3}\alpha_{3}^*)+G_{ab}^2(-\alpha_{1}\alpha_{2}-\alpha_{1}^*\alpha_{2}^*+G_{bc}^2\alpha_{3}^*(-\alpha_{1}+\alpha_{1}^*)\alpha_{2}\alpha_{2}^*\right.\\&\left.+(\mathcal{X}_{c}^2\alpha_{1}\alpha_{2}\alpha_{3}^*+\mathcal{X}_{c}^2\alpha_{1}^*\alpha_{2}^*\alpha_{3}^*+G_{bc}^2(1+2\text{i}\mathcal{X}_c\alpha_{3}^*)(\alpha_{1}-\alpha_{1}^*)\alpha_{2}\alpha_{2}^*)\alpha_{3})\right].
\end{align*}
\end{widetext}
Here, we have
\begin{equation*}\label{alpha_1,2,3}
    \begin{aligned}
        \alpha_{1}:=&1/\left(\kappa_a +\text{i}(\Delta_{a}-\delta)\right),\\
        \alpha_{2}:=&1/\left(\gamma_b +\text{i}(\Delta_{b}^\prime-\delta)\right),\\
        \alpha_{3}:=&1/\left(\gamma_c+\text{i}(\Delta_{c}^\prime-\delta)\right).
    \end{aligned}
\end{equation*}

Next, in order to study the characteristics spectra of the probe field, we deploy the standard input-output relation method~\cite{Agarwal2010,Walls1994}. That is, the cavity input field $\mathcal{E}_{\text {in }}(t)$ and the output field $\mathcal{E}_{\text {out}}(t)$ are related to each other via
\begin{equation}
\mathcal{E}_{\text {out }}(t)+\mathcal{E}_{\text {in }}(t)=2 \kappa_a a(t),
\end{equation}
which might be put into the form
\begin{equation}
\label{eq:eout1}
\mathcal{E}_{\text {out }}(t)+\mathcal{E}_{\mathrm{p}} e^{-\text{i} \Delta_p t}+\mathcal{E}_{\mathrm{l}}=2 \kappa_a \left(a_{-} e^{-\text{i} \Delta_p t}+a_{+} e^{\text{i} \Delta_p t}\right),
\end{equation}
where
\begin{equation}
\label{eq:eout2}
\mathcal{E}_{\text {out }}(t)=\mathcal{E}_{\text {out }}^{+} e^{-\text{i} \Delta_p} t+\mathcal{E}_{\text {out }}^{-} e^{\text{i} \Delta_p t}.
\end{equation}
By solving Eqs.~(\ref{eq:eout1}) and (\ref{eq:eout2}), we obtain
% \begin{equation}
% \mathcal{E}_{\text {out }}^{-}=\frac{2 \kappa_a a_{-}}{\mathcal{E}_\text{p}}-1,
% \end{equation}
% or 
\begin{equation}
\label{eq:efinal}
\mathcal{E}_{\text{out}}^{-}+1=\frac{2 \kappa_a a_{-}}{\mathcal{E}_\text{p}} := \mathcal{E}_{\text{T}}.
\end{equation}
The above relation can be obtained with the help of the homodyne technique~\cite{Walls1994}. Here, $\mathcal{E}_{\text{T}}$ has real and imaginary parts given by
\begin{equation}
\label{eq:realpart}
u_\text{p}=\frac{2\kappa_a \left(a_{-}+a_{-}^{*}\right)}{\mathcal{E}_\text{p}},  
\end{equation}
and
\begin{equation}
\label{eq:imagpart}
v_\text{p}=\frac{2 \kappa_a\left(a_{-}-a_{-}^{*}\right)}{ \mathcal{E}_\text{p}},
\end{equation}
respectively. The real part $u_\text{p}$ defines absorption, while the imaginary part $v_\text{p}$ characterizes the dispersion profile of the probe field. Similarly, we we may write the phase dispersion of the outgoing probe field as
\begin{equation}
\Phi_{t}\left(\Delta_p\right)=\arg \left[\mathcal{E}_{\text{T}}\left(\Delta_p\right)\right],
\end{equation}
which can cause transmission group delay in the vicinity of a narrow transparency window. In this way, the transmission group delay could be estimated with the help of the formula
\begin{equation}
\label{eq:groupdelay}
\tau_{g}=\frac{\text{d} \Phi_{t}\left(\Delta_p\right)}{\text{d} \Delta_p}=\frac{\text{d}\left\{\arg \left[\mathcal{E}_{\text{T}}\left(\Delta_p\right)\right]\right\}}{\text{d} \Delta_p}.
\end{equation}
Depending on the sign of $\tau_{g}$ one may determine the temporal delay profile genre, with positive and negative signs corresponding to slow and fast light propagation, respectively.

After presenting the theoretical model of the proposed cavity megnonic system, including its dynamical equations of motion, and a linearized reduced version of these equations (linearized the QLEs), we move next to a presentation of our key findings corroborated by various discussions of the system.

\section{Results and Discussions}
\label{Sec:results}
In this section, we present the main findings obtained by our model of the proposed cavity magnonic system composed of a YIG sphere placed inside the microcavity. For performing the numerical calculations, we have opted for a choice of the relevant empirical parameters based on recent experimental works~\cite{Wang2019,Wu2021}. All parameters are normalized with respect to $\omega=2\pi\times 18.6$ MHz. The remaining parameters are $\kappa_a=0.78\omega$, $\gamma_b=0.13\omega$, $\gamma_c=0.25\omega$, $\Delta_a=0.53\omega$, $\Delta_b^\prime=-0.07\omega$, $\Delta_{c}^\prime=-0.27\omega$, $G_{ab}=2.15\omega$, $G_{bc}=0.53\omega$, $\mathcal{X}_b=0.07\omega$ and $\mathcal{X}_c=0.16\omega$. In what follows, we provide results pertinent to observation of MIT and the associated dynamics of group delay obtained by properly adjusting the values of the relevant experimentally-accessible control parameters.
\begin{figure*}
\centering
	\includegraphics[width=0.7\linewidth]{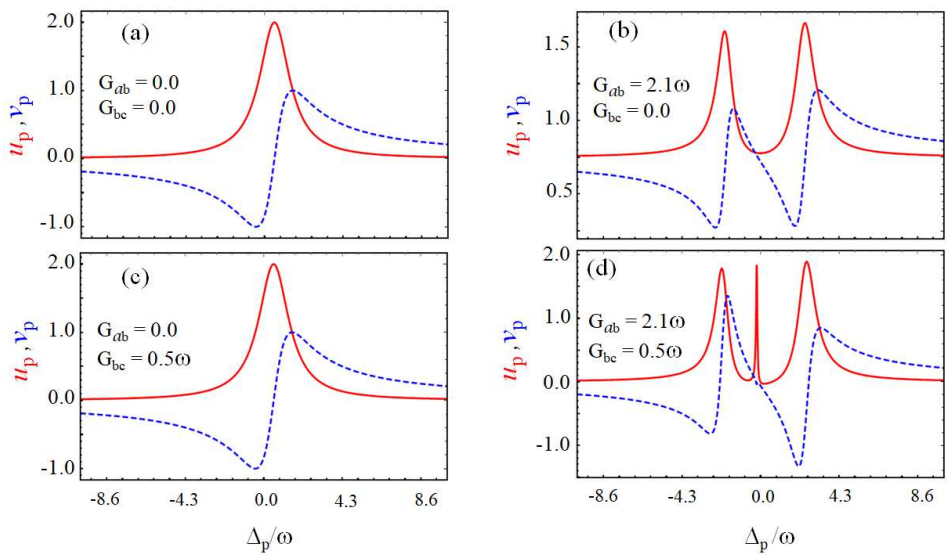}
\caption{The absorption $u_p$ and dispersion $v_p$ of the probe field displayed versus normalized optical detuning $\Delta_p/\omega$: (a) $G_{ab}=G_{bc}=0$, (b) $G_{ab}=2.1\omega$, $G_{bc}=0$, (c) $G_{ab}=0$, $G_{bc}=0.5\omega$ and (d) $G_{ab}=2.1\omega$, $G_{bc}=0.5\omega$. The remaining parameters are the same as in Sec.~\ref{Sec:results}.}
	\label{fig:two}
\end{figure*}
\subsection{Magnon-induced transparency/Absorption (MIT/MIA): Effect of coupling strengths}
Here, we examine the absorption and dispersive properties of an outgoing probe field capable of exhibiting the MIT/MIA phenomenon.
The properties of the outgoing probe field are obtained by calculating \ref{eq:realpart} and \ref{eq:imagpart}, where the real and imaginary components represent the probe field's absorption and dispersion, respectively.
Figure (\ref{fig:two}) displays the absorption $u_p$ and dispersion $v_p$ profiles of the outgoing probe field displayed as functions of normalized optical detuning $\Delta_p/\omega$, where the absorption and dispersion profiles are represented by red-solid and blue-dashed curves, respectively. These results have been obtained under different choices of the cavity-Kittle mode coupling strength $G_{ab}$ and magnon-magnon mode coupling strength $G_{bc}$ parameters (cross-Kerr coefficient). 
\begin{figure*}
\centering
	\includegraphics[width=0.7\linewidth]{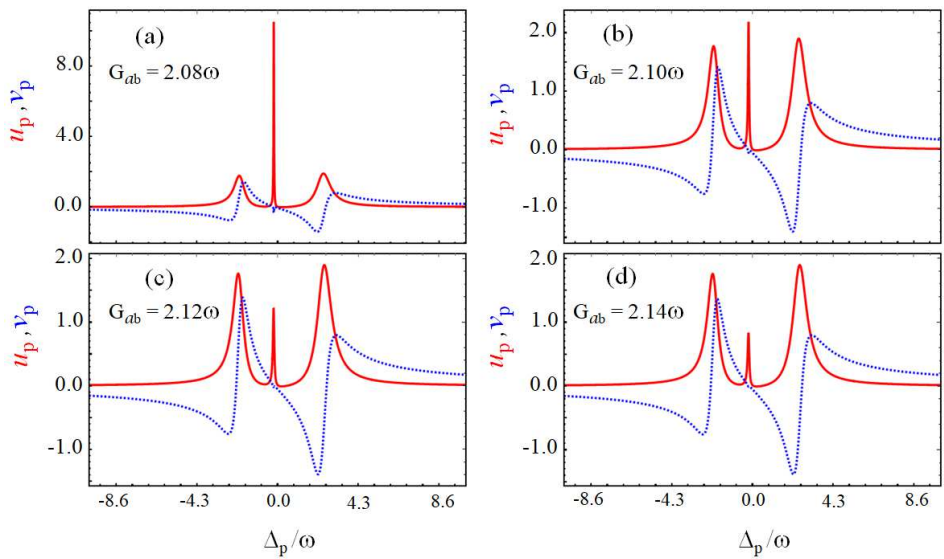}
\caption{The absorption $u_p$ and dispersion $v_p$ of the probe field versus normalized optical detuning $\Delta_p/\omega$: (a) $G_{ab}=2.08\omega$, (b) $G_{ab}=2.10\omega$, (c) $G_{ab}=2.12\omega$ and (d) $G_{ab}=2.14\omega$. The remaining parameters are as in Fig.~\ref{fig:two}.}
	\label{fig:three}
\end{figure*}
First, we consider the scenario where both coupling strengths are zero, i.e., $G_{ab}=G_{bc}=0$, as illustrated in Fig.~\ref{fig:two}(a).
In this instance, the probe field has been totally absorbed while a single MIA peak is observed. Indeed, from a basic physical viewpoint, when there is no interaction between the cavity mode and any magnon mode, there is no quantum interference between the fields and hence the applied probe field is completely absorbed. In the second scenario, i.e., for $G_{ab}=2.1\omega$ and $G_{bc}=0$, the cavity mode is coupled with only the Kittle mode while the latter is no longer coupled with the MS mode. In such case we observe a single-MIT window, see Fig.~\ref{fig:two}(b). The MIT window is formed as a result of destructive interference between the interacting fields, with a transparency width that can be enhanced by increasing the amplitude of $G_{ab}$.
Next, we analyze a different scenario in which the cavity-Kittle modes coupling strength is set equal to zero ($G_{ab}=0)$, but while the cross-Kerr coefficient is maintained at the nonzero value $G_{bc}=0.5\omega$.
Again, as in the previous case of Fig.~\ref{fig:two}(a), we observe a single MIA peak while the MIT window is found to have vanished, see Fig.~\ref{fig:two}(c). This is expected since in the absence of a cavity field, field interference phenomena cannot take place and hence MIT windows are not forthcoming. 
Furthermore, as shown in Fig.~\ref{fig:two}(d), for nonzero values of both coupling strengths, namely $G_{ab}=2.1\omega$ and $G_{bc}=0.5\omega$, we report a double-MIT window and a sharp MIA peak profile.
In other words, the MIT single window depicted in Fig.~\ref{fig:two}(b) is split into double-window profile by exploiting the extra degree of freedom added to the system. Moreover, the corresponding dispersion profile is illustrated by the blue dashed curves in Fig.~\ref{fig:two}(a-d). It illustrates how within a regime dominated by quantum interference, the coupling strength changes the dispersive behavior from anomalous to normal.
Next, we investigate how the coupling strength of the cavity-Kittle modes impacts the profile of the MIT windows.
Since prior results showed that the coupling constant $G_{ab}$ induces a single-MIT window in the absence of a cross-Kerr coefficient (see Fig.~\ref{fig:two}(b)), which then splits into a two-window profile if the cross-Kerr coefficient is also considered, see Fig.~\ref{fig:two}(d).

Figure~\ref{fig:three} depicts the absorption and dispersion profiles of the outgoing probe field plotted as a function of optical detuning for various values of cavity-Kittle mode coupling strength. For $G_{ab}=2.08\omega$ we observe a double-MIT profile whose two windows are separated by a high MIA peak realized with $\delta<\omega$, see Fig.~\ref{fig:three}(a). Moreover, by increasing the magnitude of the cavity-Kittle modes coupling strength one can further reduce the amplitude of the middle MIA peak without affecting the magnitudes of other two MIA symmetric peaks, see the example given in Fig.~\ref{fig:three}(b-d). This suggests that increasing the cavity-Kittle modes interaction strength reduces the magnitude of the MIA peak, with the ability to reach a threshold after which any further increase in this control parameter may transform the double-MIT windows into single window. Since we have previously demonstrated that in the presence of a cross-Kerr coefficient we a double-MIT windows profile can be obtained, it therefore follows that under the condition $G_{ab}\gg G_{bc}$ the effect of cross-Kerr coefficient is suppressed and hence we may obtain a single-MIT window profile.
(Note that the dispersion profile in the Figure is represented in each panel by a blue dashed line with a small kink-like peak at $\delta\approx 0$, which is not visible due to scaling.)
\begin{figure*}
\centering
	\includegraphics[width=0.7\linewidth]{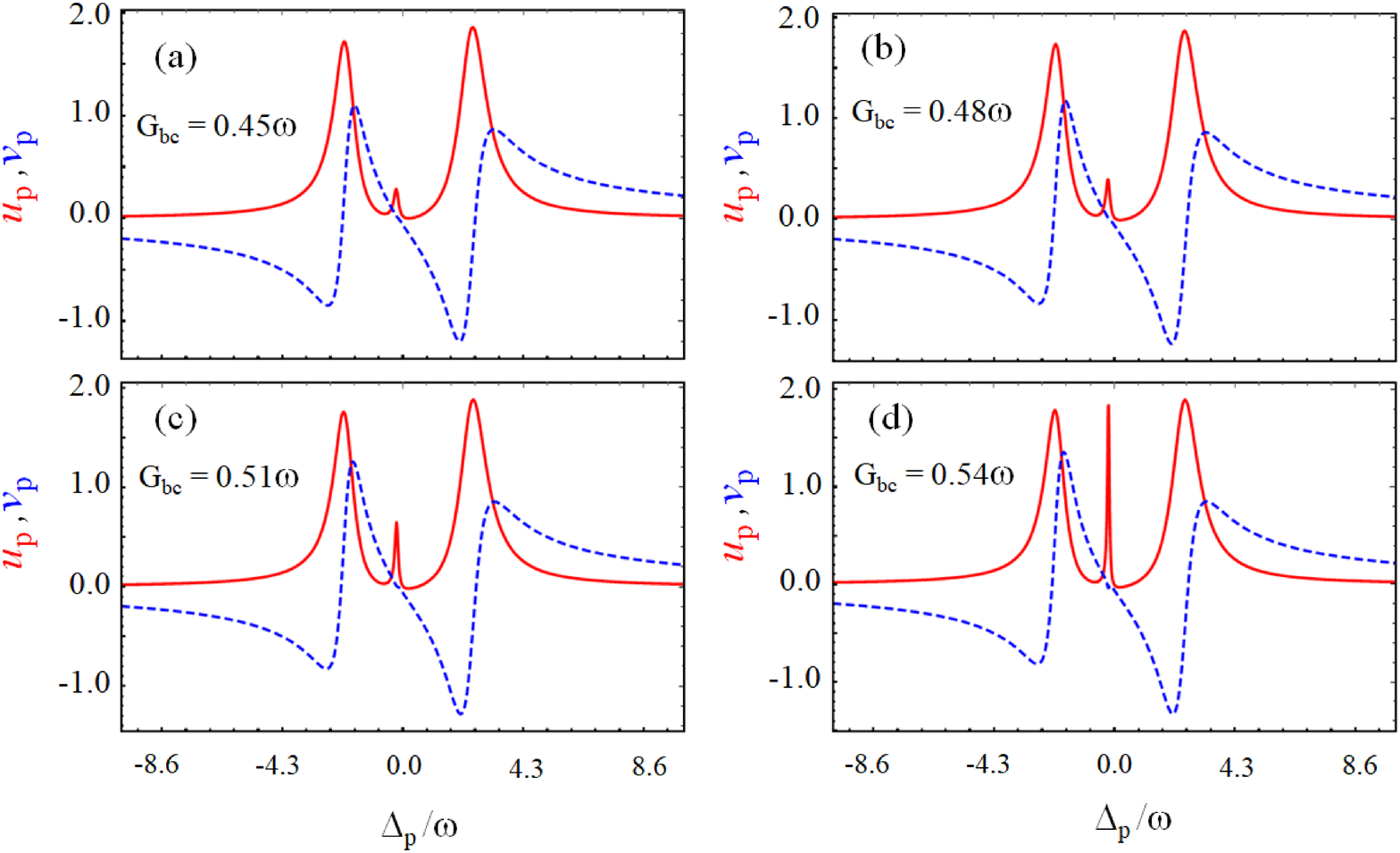}
\caption{The absorption $u_p$ and dispersion $v_p$ of the probe field versus normalized optical detuning $\Delta_p/\omega$: (a) $G_{bc}=0.45\omega$, (b) $G_{bc}=0.48\omega$, (c) $G_{bc}=0.51\omega$ and (d) $G_{bc}=0.54\omega$. The remaining parameters are as in Fig.~\ref{fig:two}.}
	\label{fig:four}
\end{figure*}

Furthermore, we analyze the effect of the cross-Kerr coefficient on the outgoing probe field. Figure~\ref{fig:four} displays the absorption $u_p$ and dispersion $v_p$ spectra of the probe field computed as a function of the optical detuning parameter $\Delta_p/\omega$.
In this case, we obtained results for various cross-Kerr coefficient $G_{bc}$ values under constant cavity-Kittle mode coupling strength $G_{ab}=2.15\omega$.
Recall that in the earlier results we have already demonstrated that in the absence of cross-Kerr coefficient, a single-MIT window (see Fig.~\ref{fig:two}(b)) can be observed, which then splits into double-MIT windows for nonzero value of cross-Kerr coefficient, as shown in Fig.~\ref{fig:two}(d). Here, we provide additional analysis of the process of the change in the probe field properties caused by increasing the cross-Kerr coefficient value.
The results obtained by our model show that as the value of the cross-Kerr coefficient $G_{bc}$ is increased, the double-MIT profile windows and the MIA profile peaks become more visible and sharply distinguishable, see Fig.~\ref{fig:four}.
Thus, we highlight the potential of utilizing proper values for the cross-Kerr coefficients in order to achieve the double-MIT profile phenomenon as noted above. In addition, the corresponding results of the dispersion profile of the outgoing probe field are shown in each panel of Fig.~\ref{fig:four}.

To recap the results of this subsection, the absorption and dispersion spectra of the probe field were illustrated, and the single-to-double MIT windows and MIA profile were found to be realizable by proper tuning of relevant control parameters, specifically the cavity-Kittle modes and cross-Kerr coefficients. We observed that for a large value of the cavity-Kittle modes coupling constant, the cross-Kerr effects are suppressed, and the double-MIT window profile is switched to a single-MIT window profile, after which the effect of self-Kerr coefficients on the MIT windows and MIA peaks were investigated. 
\begin{figure}
	\includegraphics[width=0.85\linewidth]{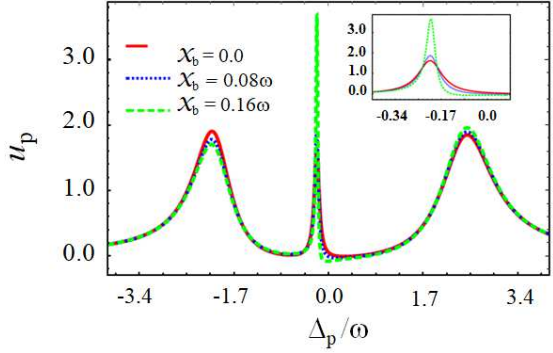}
\caption{The absorption $u_p$ profile of the probe field versus normalized optical detuning $\Delta_p/\omega$ for: $\mathcal{X}_b=0$ (red-solid curve), $\mathcal{X}_b=0.08\omega$ (blue dotted curve), and $\mathcal{X}_b=0.16\omega$ (green dashed curve). The inset shows the more broader view of the absorption profile of the outgoing probe field. The additional parameters are the same as in Fig.~\ref{fig:two}.}
	\label{fig:five}
\end{figure}
\subsection{Magnon-induced transparency/Absorption (MIT/MIA): Effect of self-Kerr coefficients}
In this subsection, we investigate the impact of the self-Kerr coefficient on the MIT and MIA phenomena. Figure~\ref{fig:five} depicts the absorption profile $u_p$ of the outgoing probe field plotted as a function of normalized optical detuning $\Delta_p/\omega$ for different values of self-Kerr coefficient of Kittle mode $\mathcal{X}_b=\Delta_p/\omega$ (red-solid curve), $\mathcal{X}_b=0.08\omega$ (blue-dotted curve) and $\mathcal{X}_b=0.16\omega$ (green-dashed curve), while the self-Kerr coefficient of the MS mode remains fixed, i.e., $\mathcal{X}_c=0.16\omega$. For $\mathcal{X}_b=0$, we observe the profile of double-MIT windows separated by high MIA peak. The MIA peaks show the asymmetric behaviour. A further increase in the self-Kerr coefficient of the Kittle mode can reduce the amplitude of the MIA peak that separate the double-MIT windows while also causing an enhancement of the asymmetric behaviour of the other two MIA peaks. The inset in Figure~\ref{fig:five} displays the enhanced estimation of the MIA profile of the probe field against the various values of the self-Kerr coefficient of the Kittle mode. Therefore, in this case the self-Kerr coefficient of Kittle mode can only influence the asymmetric behavior of the MIA profile of the outgoing probe field with no impact on the amplitude of the MIT windows.  
\begin{figure}
	\includegraphics[width=0.85\linewidth]{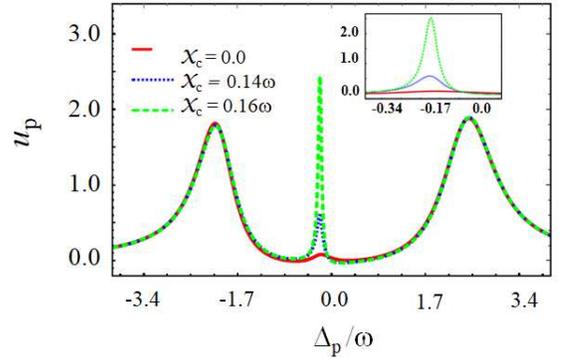}
\caption{The absorption $u_p$ profile of the probe field versus normalized optical detuning $\Delta_p/\omega$ for: $\mathcal{X}_c=0$ (red-solid curve), $\mathcal{X}_c=0.14\omega$ (blue dotted curve), and $\mathcal{X}_c=0.16\omega$ (green dashed curve). The inset depicts the broader view of the absorption profile of the probe field. The remaining parameters are as in Fig.~\ref{fig:two}.}
	\label{fig:six}
\end{figure}

Next, we investigate how changing the self-Kerr coefficient of the MS mode can modify the MIT and MIA peaks profile. Fig.~\ref{fig:six} depicts the absorption profile $u_p$ of the outgoing probe field plotted as a function of normalized optical detuning $\Delta_p/\omega$ for different values of the self-Kerr coefficient of the MS mode. The result are shown for $\mathcal{X}_c=0$ (red-solid curve), $\mathcal{X}_c=0.14\omega$ (blue-dotted curve) and $\mathcal{X}_c=0.16\omega$ (green-dashed curve), whereas the self-Kerr coefficient of the Kittle mode is kept fixed at $\mathcal{X}_b=0.07\omega$. The results illustrate that the absorption profile of the outgoing probe field displays a wider single-MIT window in the absence of self-Kerr coefficient of MS modes, i.e., $\mathcal{X}_c=0$. However, when by increasing the value of $\mathcal{X}_c$, each single-MIT profile windows split into a double window. Further increase in this control parameter may result in a more enhanced visibility of the double-MIT windows profile, a behavior illustrated in the inset of in Fig.~\ref{fig:six}. Thus, the self-Kerr coefficient of the MS mode provides a new degree of freedom capable of splitting single-MIT window profiles into double window profiles.

To summarize, in this subsection, we examined the impact of the self-Kerr coefficients of both spin modes on the MIT and MIA spectra, where it was demonstrated that the self-Kerr coefficient of the Kittle mode is capable of modifying the asymmetric behavior of the MIA profile, whereas the self-Kerr coefficient of the MS mode is responsible for splitting the single-MIT into two distinct windows. Thus, investigating the impact of self-Kerr coefficients of the Kittle and MS modes on the MIT windows profile is significant for the correct interpretation of interference phenomena in cavity magnonics. In the following subsection, we explore the propagation of the group delay of the outgoing probe field, which exhibits slow and fast light phenomena.
\begin{figure}
	\includegraphics[width=0.8\linewidth]{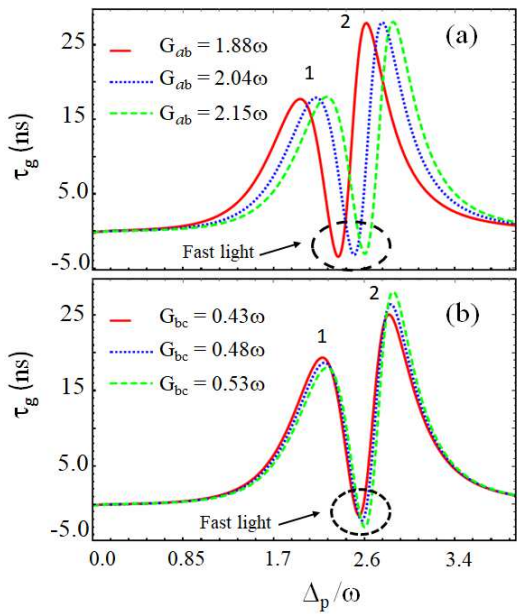}
\caption{Group delay $\tau_\text{g}$ versus normalized optical detuning $\Delta_p/\omega$ for: (a) $G_\text{ab}=1.88\omega$ (red-solid curve), $G_\text{ab}=2.04\omega$ (blue-dotted curve) and $G_\text{ab}=2.15\omega$ (green-dashed curve), and (b) $G_\text{bc}=0.43\omega$ (red-solid curve), $G_\text{bc}=0.48\omega$ (blue-dotted curve) and $G_\text{bc}=0.53\omega$ (green-dashed curve), where $G_\text{ab}$ and $G_\text{bc}$ are the coupling constants corresponding to cavity-Kittle modes and magnon-magnon modes. The region $\tau_\text{g} < 0$ enclosed by circle signifies the fast light regime, whereas the two pointed peaks represent the slow light scenario. The remaining parameters are the same as in Fig.~\ref{fig:two}.}
	\label{fig:seven}
\end{figure}
\subsection{Dynamics of slow and fast light}
The phenomenon of slow and fast light has been observed in atomic vapors and solid-state materials using a number of techniques. Controlling the group velocity of light pulses such that to cause very slow or very fast propagation is one prominent application of these approaches~\cite{Boyd2009,Scully2003}. For instance, processes of EIT in atomic vapors or Bose-Einstein condensate were employed in slow light studies~\cite{Kasapi1995,Kien2009}. Hau et al. presented an experimental demonstration of EIT in an ultracold sodium atom gas, where optical pulses move at a speed 20 million times slower than the speed of light in a vacuum. Aside from slow light, fast light was reported in atomic cesium gas and a silicon microphotonic device~\cite{Hau1999}. Afterwards, Safavi-Naeini et al. demonstrated the ability of designed photon-phonon interactions to control the velocity of light while also exhibiting EIT and programmable optical delays in a nanoscale optomechanical crystal~\cite{Safavi2011}. Therefore, it is of natural interest to investigate whether there is a viable physical configuration capable of switching from slow to fast light or vice versa. In the following, we present a cavity magnonics system wherein we investigate the propagation of group delay and address the switching from slow to fast light within a single configuration. 

The results for slow and fast light were obtained by using the mathematical expression for group delay Eq.~\ref{eq:groupdelay}. Fig.~\ref{fig:seven} illustrates the propagation of group delay $\tau_\text{g}$ plotted as a function of normalized optical detuning $\Delta_p/\omega$ for different values of: (a) $G_\text{ab}=\{1.88\omega,~2.04\omega,~2.15\omega\}$ and (b) $G_\text{bc}=\{0.43\omega,~0.48\omega,~0.53\omega\}$, where $G_\text{ab}$ and $G_\text{bc}$ denote the cavity-Kittle modes coupling strength and cross-Kerr coefficient, respectively. Here, we demonstrated the results of the group delay in the region where the double-MIT window profiles exist. As can be seen from Fig.~\ref{fig:three} and Fig.~\ref{fig:four}, both MIT windows dips occur on the left of the optical detuning, i.e., $\Delta_p/\omega < 0$, because we employ off-detuning values of the effective detuning values of Kittle and MS modes, i.e.,  $\Delta_b^\prime=-0.07\omega$, $\Delta_{c}^\prime=-0.27\omega$. Initially, we present the results of the group delay versus optical detuning for various values of the cavity-Kittle modes coupling constant $G_\text{ab}$, see Fig.~\ref{fig:seven}. The results also show that increasing the cavity-Kittle modes coupling constant reduces the amplitude of the dip at $\Delta_\text{p}  \approx 2.6\omega$. Because the group delay value at this point is negative, this corresponds to fast light. These findings are in line with the prior results given in Fig.~\ref{fig:three}, where the cavity-Kittle modes coupling constant lowered the MIA profile of the probe field and had an effect on the amplitude of MIT windows. Physically, the group delay is negative, indicating slow light during the MIT windows caused by destructive interference of fields and vice versa.
To further investigate the impact of cross-Kerr effect on the propagation of group delay, Fig.~\ref{fig:seven}(b) shows the propagation of group delay vs. optical detuning for different values of cross-Kerr coefficients $G_\text{bc}$. The results are shown in the range of optical detuning where the two MIT windows exist, as demonstrated in Fig.~\ref{fig:four}. The group delay has two distinct peaks that represent slow light dynamics, however the dip enclosed by a dashed circle represents fast light dynamics (because we noticed in previous results (see Fig.~\ref{fig:four}) that increasing the value of cross-Kerr coefficient changes the MIA profile of the probe field), hence illustrating the slow and fast light effects in the two MIT windows profiles. Therefore, the proposed scheme suggests a mechanism for switching from slow to fast light in a single configuration. 

\section{Conclusions}
\label{Sec:conclusion}
We investigated various optomechanical nonlinear effects involving cross- and self-Kerr interactions in a cavity magnonic system composed of a microcavity and a ferromagnetic material (yttrium iron garnet sphere) that exhibits both Kittle and magnetostatic modes. Based on our analytical and numerical results, we observed magnon-induced transparency (MIT) and magnon-induced absorption (MIA) caused by quantum interference of optical fields inside a cavity. The impact of coupling parameters such as cavity-Kittle modes and magnon-magnon modes (cross-Kerr effect) is investigated. In addition, we explained how to properly adjust these parameters in order to achieve single-to-double MIT windows. Furthermore, it was established that the self-Kerr effect of both Kittle and magnetostatic (MS) modes can alter the MIT and MIA phenomena, with the self-Kerr effect of Kittle mode causing asymmetric MIA behavior and the self-Kerr effect of MS mode splitting the single-MIT window into two distinct windows. The propagation of group delay was investigated in the vicinity of MIT windows. It was observed that increasing the cross-Kerr effect can help maintain slow light for a longer period of time. Our theoretical model could provide a new platform for studying nonlinear effects in cavity magnonics, which have applications in quantum memory~\cite{Fiore2011}, quantum entanglement~\cite{Li2018}, and quantum information processing.

% \section*{ACKNOWLEDGMENTS}
% AM acknowledges the support of postdoctoral fund of ZJUI.  
\section*{Disclosures}
The authors declare no conflicts of interest. 
\section*{Data availability.}
No data was used for producing this work.
%%%%%%%%%%%%%%%%%%%%%%% References 
%\bibliography{biblio}% Produces the bibliography via BibTeX.
%

\end{document}